    \documentstyle[proceedings,numreferences,epsfig]{crckapb}
\newcommand{\rk}{\mbox{\boldmath $k$}}
\newcommand{\rkp}{\mbox{\boldmath $k^{\prime}$}}

\def\pom{{I\!\!P}}

\begin{opening}
\title{The QCD gluon ladders and HERA structure function}

\author{A. Lengyel $^1$}
\author{M.V.T. Machado $^2$}

\institute{$^1$ Institute of Electron Physics, National
Academy of Sciences of Ukraine. Universitetska 21,
UA-88016 Uzhgorod, Ukraine.\\
$^2$ High Energy Physics Phenomenology Group, IF-UFRGS.\\
Caixa Postal 15051,
CEP 91501-970, Porto Alegre, RS, Brazil.}

\end{opening}

\runningtitle{The QCD gluon ladders and HERA structure function}

\begin{document}


\begin{abstract}
We report on the extension of the data fitting considering
  the QCD inspired model based on the summation of gluon
  ladders applied to the $ep$ scattering. In lines of a
  two Pomeron approach, the structure function $F_2$ has
  a  hard piece given by the  model and the remaining soft
  contribution: a soft Pomeron  and  non-singlet content.
  In this paper, we carefully estimate the relative
  role of the hard and the soft pieces from a  global fit
  in a  large span of $x$ and $Q^2$.
\end{abstract}

\section{Introduction}

The HERA small-$x$  data \cite{MKlein} have introduced a challenge to the phenomenologists in order to describe the strong growth of the inclusive structure function $F_2$ as the Bjorken scale $x$ decreases, supplemented by the  scaling violations on  the hard scale given by the photon virtuality $Q^2$. Concerning the Regge approach, at high energy the $ep$ scattering process is dominated by the exchange of the Pomeron trajectory in the $t$-channel. From the hadronic phenomenology, this implies that the structure function would present a mild increasing on energy ($s\simeq Q^2/x$) since the soft Pomeron intercept ranges around $\alpha_{\pom}(0)\approx 1.08$. Such behavior is in contrast with high energy $ep$ data, where the effective intercept takes values $\lambda_{\mathrm{eff}}\simeq 1.3-1.4$. In the Regge language this situation can be solved by introducing the idea of  new poles in the complex angular momentum  plane, for instance rendered in the multipoles models \cite{dippom,DM,Cudell}, producing quite successful data description. Other proposition  is the two Pomeron model \cite{DL}, introducing  an additional hard intercept and corresponding residue. However, a shortcoming from these approaches is the poor knowledge about the behavior on virtuality, in general modeled in an empirical way through the vertex functions.

On the other hand, the high photon virtuality allows applicability of the QCD perturbative methods. The DGLAP formalism \cite{DGLAP} is quite successful in describing most of the measurements on structure functions at HERA and hard processes in the hadronic colliders. This feature is even intriguing, since its theoretical limitations at high energy are well known \cite{OPEbreak}. Other perturbative approach is the BFKL formalism \cite{LOBFKL}, well established at LO level but not yet completely understood at NLO accuracy. The main issue in the NLO BFKL effects  is the correct account of the sub-leading corrections in the  all orders resummation \cite{Salam}. The LO BFKL approach can describe HERA structure function in a limited kinematical range, i.e. at not so large $Q^2$ and small-$x$. A consistent treatment considering higher order resummations is currently being available and applications should be allowed  in a  near future.

In this contribution we report on the extension of the data fitting to the HERA structure function $F_2$ using as a model for the hard Pomeron the  finite sum of gluon ladders \cite{Trunkbfkl}. The model  is based on the truncation of the BFKL series considering only the  first few orders in the strong coupling perturbative expansion, where subleading contributions can be absorbed in the adjustable  parameters. From the phenomenology on hadronic collisions \cite{Fiore}, just  three orders, $\sim [\alpha_s\ln (1/x)]^2$,  are  enough to describe current accelerator data. The hard Pomeron model should be supplemented  by a  soft piece accounting for the non-perturbative contributions to the process. The description therefore turns out similar to the two Pomeron model \cite{DL}, with the advantage of a complete knowledge of the  behaviors on $x$ and $Q^2$. The original model contains a reduced number of adjustable parameters: the normalization ${\cal N}_p$ and  the non-perturbative scale $\mu^2$ from the proton impact factor and the parameter $x_0$ scaling the logarithms on energy.

In spirit of a global analysis, in the Ref. \cite{PLB2002} two distinct choices  for the soft Pomeron were analised. The resulting two Pomeron model was successful in describing  data on structure function $F_2$ and its derivatives (slopes on $Q^2$ and $x$) for $x \leq 2.5\,10^{-2}$  and $0.045\leq Q^2\leq 1500$ GeV$^2$. Here, we extend the description to the whole $x$ range and analise the relative role of the hard and soft pieces in the results. This work is organized as follows. In the next section we shortly review the main expressions for the hard piece given by the summation up to the two-rung ladder contribution. In   Sec. (3), an overall fit to the recent deep inelastic data is performed based on the hard contribution referred above supplemented by the remaining  soft Pomeron and non-singlet contributions. In the last section we draw up our conclusions.

\section{The hard contribution: summing gluon ladders}

Here we review the elements needed to calculate the structure function using the finite sum of gluon ladders in the $ep$ collision, with center of mass energy $W$.  The proton inclusive
structure function, written in terms of the cross sections for the
scattering of transverse or longitudinal polarized photons, reads
as \cite{Predazzibook}
\begin{eqnarray} F_2(x,Q^2) &=& \frac{Q^2}{4\pi^2 \alpha_{\rm{em}}}
\left[\sigma_T(x,Q^2) + \sigma_L(x,Q^2)\right] \,,\\
\sigma_{T,\,L}(x,Q^2)& = & \frac{{\cal G}}{(2\pi)^4} \, \int
\frac{d^2\rk}{\rk^2}\,\frac{d^2\rkp}{\rkp^2}\,\Phi^{\gamma^*}_{T,\,L}(\rk)\,
F(x,\rk,\rkp)\,\Phi_p(\rkp)\,, \label{eq1}
\end{eqnarray}
where ${\cal G}$ is the color factor for the
color singlet exchange and $\rk$ and $\rkp$ are the transverse
momenta of the exchanged reggeized gluons in the $t$-channel. The
$\Phi^{\gamma^*}_{T,\,L}(\rk)$ is the virtual photon impact
factor  and $\Phi_p(\rkp)$ is the proton
impact factor. The first one is well known in perturbation theory
at leading order, while the latter is modeled
since in the proton vertex there is no hard scale to allow
pQCD calculations. The kernel $F(x,\rk,\rkp)$ contains the dynamics of the process, for instance the BFKL kernel.

The amplitudes can be calculated order by order: for instance the Born contribution coming from  the two gluon exchange and the one-rung ladder contribution read as,
\begin{eqnarray}
{\cal A}^{(0)} & = & \frac{2\,\alpha_s \, W^2}{\pi^2}\,\sum_f \, e^2_f\,
\int \frac{d^2\rk}{\rk^4}\,\Phi^{\gamma^*}_{T,\,L} (\rk)\,\Phi_p(\rk)\,,\nonumber\\
 {\cal A}^{(1)} & = &  \frac{6 \alpha^2_s
W^2}{8 \pi^4}\, \sum_f\, e^2_f\, \ln \left( \frac{W^2}{W^2_0}\right)\, \int \frac{d^2\rk}{\rk^4}\,
\frac{d^2\rkp}{\rkp^4}\,
 \Phi^{\gamma^*}_{T,\,L}(\rk),{\cal K}(\rk,\rkp) \,\Phi_p(\rkp) \,,\nonumber
\label{eq2}
\end{eqnarray}
where $\alpha_s$ is  considered fixed since we
are in the framework of the LO BFKL approach.
The perturbative kernel  ${\cal K}(\rk,\rkp)$ can be calculated order by order in the perturbative expansion \cite{Predazzibook}.  The Pomeron is
 attached to the off-shell incoming photon through the quark loop
diagrams, where the Reggeized gluons are attached to the same and to
different quarks in the loop. The virtual photon impact factor averaged over the transverse polarizations reads as \cite{Balitsky},
\begin{eqnarray}
\Phi^{\gamma^*}_{T,\,L}(\rk)=\frac{1}{2} \int_0^1
\frac{d\tau}{2\pi} \,\int_0^1 \frac{d \rho}{2\pi}\,
\frac{\rk^2(1-2\tau\, \tau^{\prime})(1-2\,\rho \,\rho^{\prime})}{\rk^2
\,\rho\, \rho^{\prime} + Q^2 \rho \,\tau\, \tau^{\prime}},
\end{eqnarray}
where $\rho$, $\tau$ are the Sudakov variables associated
to the momenta in the photon vertex and  $\tau^{\prime}\equiv (1-\tau)$
and $\rho^{\prime}\equiv (1-\rho)$.

Gauge invariance requires  the proton impact factor vanishing at $\rk$ going to zero and is modeled  in a simple way,
\begin{eqnarray} \Phi_p(\rkp)={\cal N}_p
\,\frac{\rkp^2}{\rkp^2 + \mu^2},
\end{eqnarray}
where ${\cal N}_p$ is the unknown normalization of the proton
impact factor and $\mu^2$ is a scale which is typical of the
non-perturbative dynamics. These scales will  be considered adjustable parameters in the analysis.  Considering the electroproduction process,
summing the  first orders in perturbation theory  we can write the
expression for the inclusive structure function,
\begin{eqnarray}
F_2^{\mathrm{hard}}(x,Q^2) & = & \frac{8}{3}\,\frac{\alpha^2_s}{\pi^2} \sum_f e^2_f \, {\cal N}_p
\left[\, I^{\,(0)}(Q^2,\mu^2) + \frac{3\,\alpha_s}{\pi}\,\ln
\frac{x_0}{x}\,I^{\,(1)}(Q^2,\mu^2)\,  \right. \nonumber
\\
& &  + \left.
 \frac{1}{2}\,\left(\frac{3\,\alpha_s}{\pi}\,\ln \frac{x_0}{x}\right)^2
\,I^{\,(2)}(Q^2,\mu^2)\, \right],
\label{eq7}
 \end{eqnarray}
where the functions $I^{\,(n)}(Q^2,\mu^2)$ correspond to the $n$-rung gluon ladder contribution. The quantity  $x_0$ gives the scale normalizing  the logarithms on energy for the LLA BFKL approach, which is arbitrary and enters as an
additional parameter. The contributions are written explicitly as,
\begin{eqnarray}
I^{\,(0)} & = &  \frac{1}{2} \ln^2 \left(
\frac{Q^2}{\mu^2} \right) + \frac{7}{6}\ln \left(
\frac{Q^2}{\mu^2} \right) + \frac{77}{18} \,,\\
I^{\,(1)} & = &  \frac{1}{6} \ln^3 \left(
\frac{Q^2}{\mu^2} \right) + \frac{7}{12}\ln^2 \left(
\frac{Q^2}{\mu^2} \right) + \frac{77}{18}\ln \left(
\frac{Q^2}{\mu^2} \right) + \frac{131}{27} + 2\,\zeta (3)\,,\\
I^{\,(2)} & = & {1\over24}\ln^4 \left(\frac{Q^2}{\mu^2}\right)+
\frac{7}{36}\ln^3 \left(\frac{Q^2}{\mu^2}\right)+ \frac{77}{36}\ln^2 \left(\frac{Q^2}{\mu^2}\right) \nonumber\\
& & + \left(\frac{131}{27} +4 \zeta(3)\right) \ln \left( \frac{Q^2}{\mu^2}\right) + \frac{1396}{81} - \frac{\pi^4}{15} + \frac{14}{3}\zeta(3)\,,
\label{eq10}
\end{eqnarray}
where  $\zeta (3)\approx 1.202$.  The main result in Ref. \cite{PLB2002} is in a good agreement, in terms of a $\chi^2/\mathrm{dof}$ test,  for the
inclusive structure function in the range  $0.045 \leq Q^2 \leq 1500$ GeV$^2$, once considering the restricted kinematical constraint $x \leq 0.025$.
 The non-perturbative contribution (from the soft dynamics), initially considered as a background,  was found  to be not negligible. We have estimated that such effects introduce a correction of the same order in the overall normalization. In the next section we perform a global analysis in lines of our previous work \cite{PLB2002}, extending the range on $x$ fitted by adding the non-singlet contribution modeled through the usual Regge parameterizations.

\section{Fitting results and conclusions}

In order to perform the fitting procedure, for the hard piece one uses Eq. (\ref{eq7}) and for the soft piece we have selected a model with the  most economical number of parameters. For  this purpose it was  selected the latest
 version \cite{KMP} of the CKMT model \cite{CKMT}:

\begin{eqnarray}
F^{\mathrm{soft}}_2(x,Q^2)= A\,\left(
\frac{x_1}{x} \right)^{\Delta(Q^2)} \left(
\frac{Q^2}{Q^2+a}\right)^{\Delta(Q^2)+1}\,\left(1-x\right)^{n_s({Q^2})}
\,, \label{softpom}\\
{\Delta(Q^2)}= \Delta_{0} \left( 1+
\frac{Q^2 \Delta_{1}}{\Delta_{2}+Q^2}\right)\,,
\hspace{0.6cm}
{n_s}(Q^{2})=\frac{7}{2} \,
\left(1+\frac{Q^{2}}{Q^{2}+b}\right)\,,
\end{eqnarray}
where $\Delta(Q^2)$ is the Pomeron intercept. The non-singlet term is taken in a form,
\begin{eqnarray}
 F^{\mathrm{ns}}(x,Q^2) & = & A_R\, x^{1-\alpha_{R}}\,\left(\frac{Q^2}{Q^2+a_R}\right)^{\alpha_{R}}\,(1-x)^{n_{ns}({Q^2})}\, \label{softreg} \\
 n_{ns}(Q^2) & = & \frac{3}{2}\,\left(1+\frac{Q^2}{Q^2+d}\right).
 \end{eqnarray}

Concerning the hard piece, Eq. (\ref{eq7}), we selected the
 overall normalization factor as a free parameter, defined as ${\cal
N}=\frac{8}{3}\,
 \frac{\alpha^2_s}{\pi^2}\, \sum e^2_f \,{\cal N}_p$, considering
 four active flavors. It was also supplemented by a factor  $(1-x)^{n_{s}}$ accounting for the large $x$ effects. For the fitting procedure we consider the data set containing all
available HERA data for the proton structure function $F_2$
\cite{H1c1}, \cite{ZEUSc1} - \cite{H1c6} adding only several data set of
fixed target experiments \cite{fixed}. For the fit we have used
$900$ experimental points for all available
 $x$ and $0.045\leq Q^2 \leq 30000$ GeV$^2$.

\begin{figure}[h]
\centerline{\epsfig{file=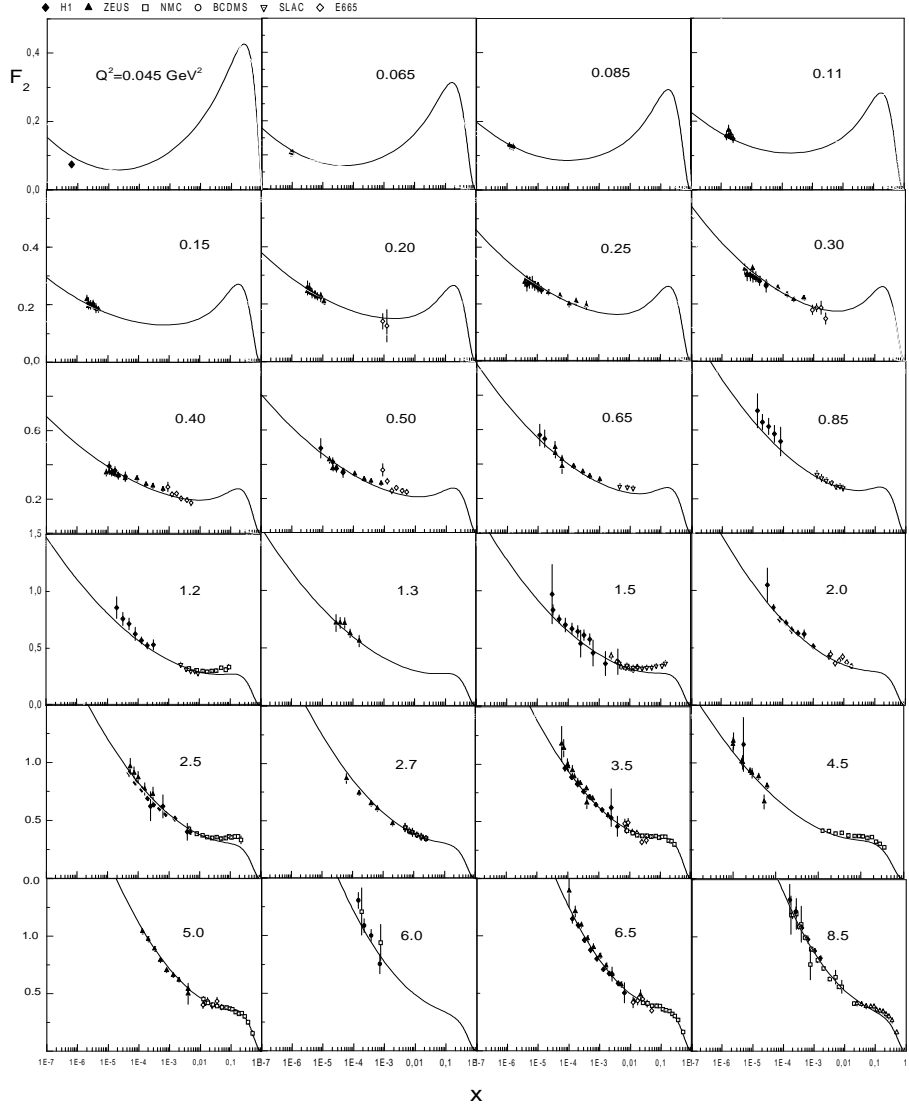,width=13cm,height=16cm}}
\caption{The inclusive structure
function at small $Q^{2}$ virtualities. The procedure (I) and (II) produce equivalent curves.}
\label{fig1}
\end{figure}

\begin{figure}[h]
\centerline{\epsfig{file=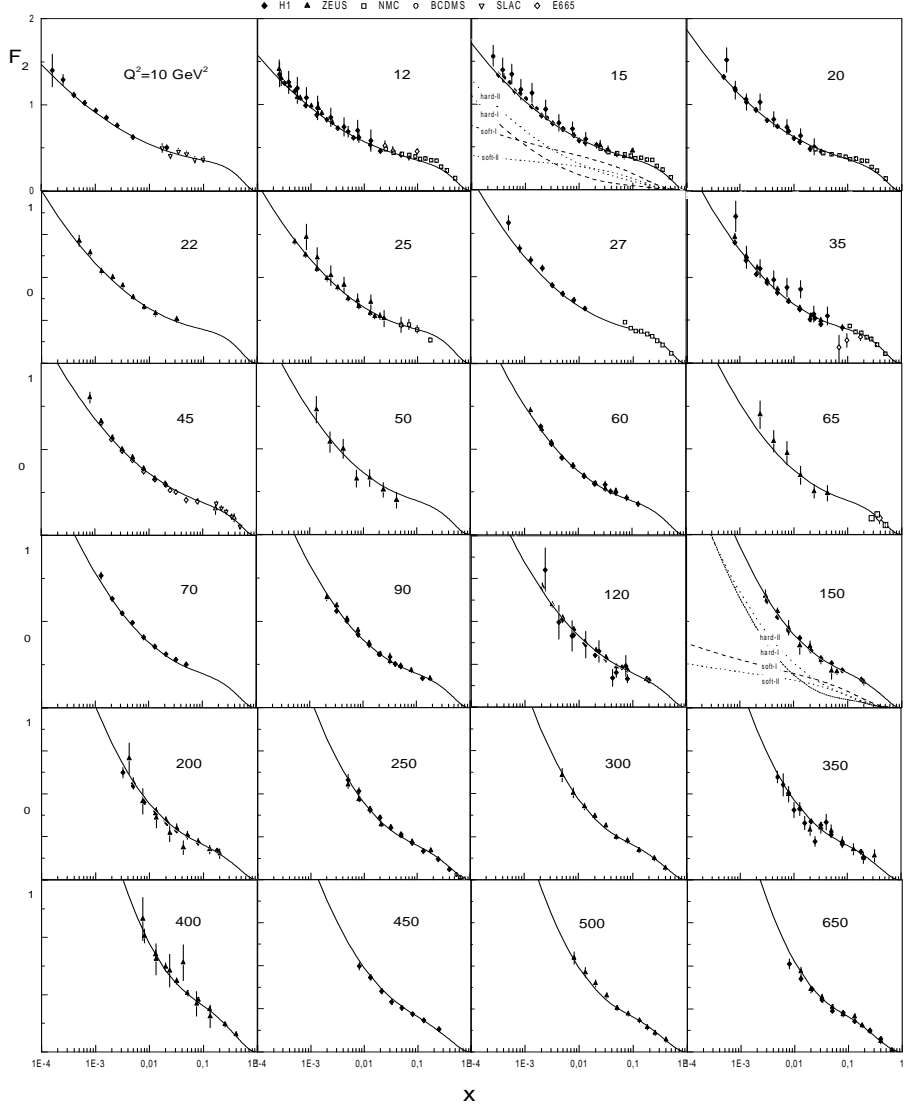,width=13cm,height=16cm}}
\caption{The results for  the inclusive structure
function at medium $Q^{2}$ virtualities.}
\label{fig2}
\end{figure}

\begin{figure}[h]
\centerline{\epsfig{file=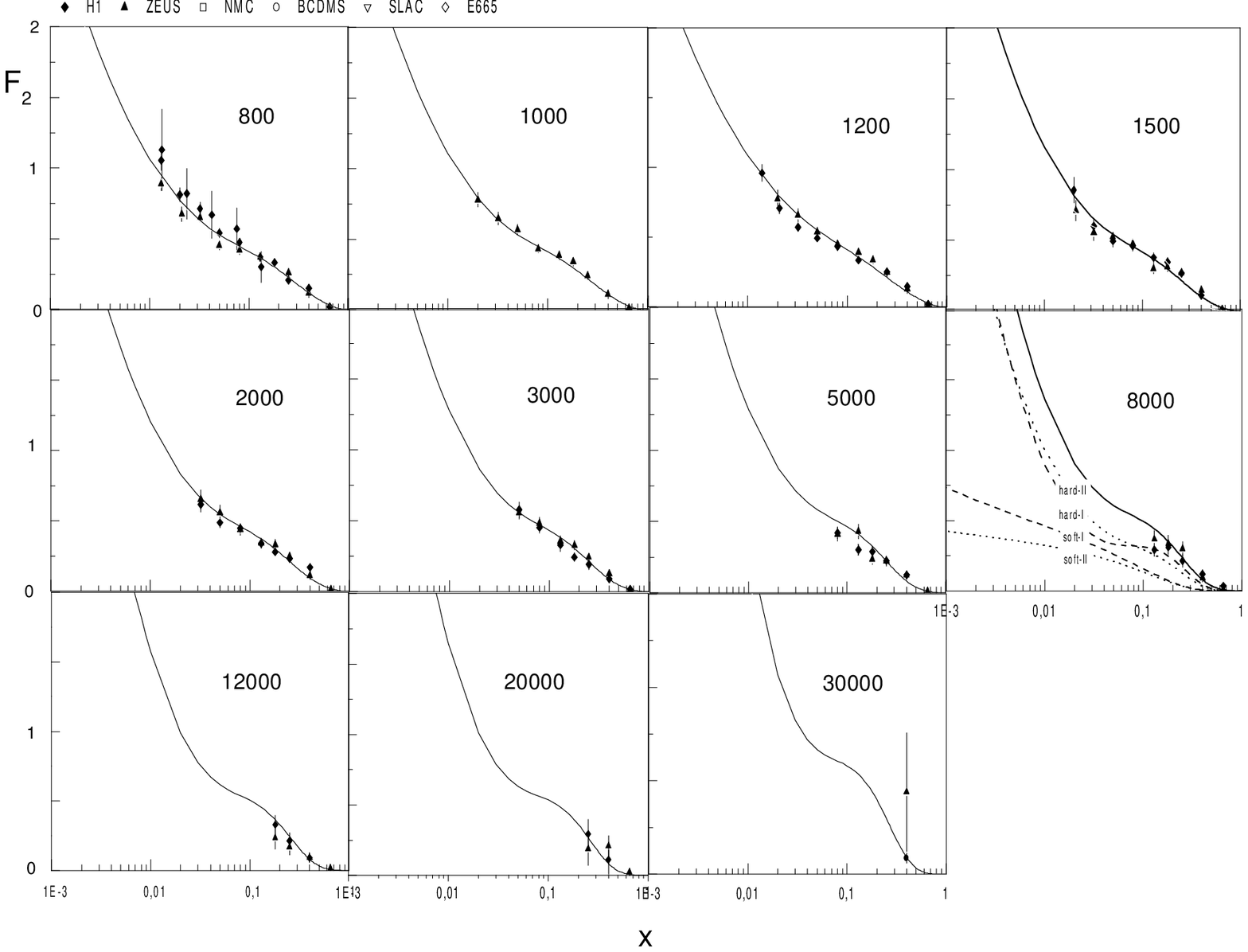,width=14cm,height=15cm}}
\caption{The results  for the inclusive structure function at
large $Q^{2}$ virtualities.}
\label{fig3}
\end{figure}

In the Figs. (\ref{fig1}) (\ref{fig2}) and (\ref{fig3}) are shown the results considering three different fitting procedures: (I) overall fit using the hard piece and the soft pieces, Eqs. (\ref{softpom}) and (\ref{softreg}); (II) overall fit restricting the soft Pomeron intercept; (III) fitting only the hard piece plus non-singlet contribution.  The best fit parameters for these procedures are presented in  Table (1). In the following we discuss each one, pointing out  the relative contribution from the hard and soft pieces.  The procedure (I) provides
the same quality description {$\chi^{2}/\mathrm{dof}=1.12$} in the
whole available interval of Bjorken variable and virtuality ($0\leq x \leq 1$ and $0.045 \leq Q^2 \leq 30000$)  as  in the previous analysis \cite{PLB2002}. It is worthy note that the relative
contribution of soft Pomeron remains  the same in comparison with this analysis, i.e. the hard and soft pieces are comparable in the considered experimental domain. This feature is related to the choice for a CKMT Pomeron, i.e. a $Q^2$ dependent intercept ranging from $1.07$ up to $1.17$. The fit is comparable with those using a two-Pomeron analysis \cite{DM,DL}.

In the procedure (II), we have restricted the soft Pomeron intercept through the following way,
\begin{eqnarray}
 {\Delta(Q^2)}= \Delta_0 \left(
\frac{Q^2}{\Delta_1+Q^2}\right) + \Delta_2\,,
\end{eqnarray}
where a relatively good result was obtained in the restricted interval of vituality $1.5 \leq Q^2 \leq 30000$ GeV$^2$. The parameters obtained are shown in Table (1). The quality is slightly worst than the procedure (I).

\begin{table}[htb]
\begin{center}
\caption{Parameters of the fit. Procedure (I): hard plus soft terms; Procedure (II): similar to analysis (I), but restricting soft Pomeron intercept; Procedure (III): only hard term and non-singlet contribution.}
\begin{tabular}{lllll}
\hline
&   &   I       &  II     & III  \\
\hline
  & $\cal N$ & 0.0235  & 0.0188 & 0.0233 \\
Hard Pomeron & $ \mu^2 $ & 1.36  & 0.374 & 0.146 \\
 & $x_0$ & 0.0077    & 0.00589 & 0.00327\\
& $\alpha_s$ & 0.2 (fixed)& 0.13(fixed)& 0.086\\
\hline
& A & 0.3085 & 0.282 & - \\
& a & 0.693  & 2.12 & -\\
Soft Pomeron & $\Delta_0$ & 0.0118 & 0.07 & -\\
& $\Delta_1$ & 13.75 & 1.13 & -\\
& $\Delta_2$ & 5.85  & 0.02 & -\\
& $x_1$ & 0.115 & 1.0   & -  \\
& b & 26.7 & 4.94 & -
\\ \hline
      & $A_R$ &  6.15  & 6.42   & 5.19  \\
Non-singlet & $\alpha_{R}$ & 0.7(fixed) & 0.7(fixed) & 0.7(fixed)\\
 & $a_{R}$ & 603 & 671 & 918\\
  & d & 0.908 & 0.575 & 0.0\\
\hline
$\chi^2/\mathrm{d.o.f.}$ &    & 1.12 & 1.34 & 1.40 \\
\hline
\end{tabular}
\end{center}
\end{table}

Finally, we  succeeded to perform a fit not considering the soft Pomeron contribution, but keeping  the non-singlet term. The procedure (III) gives reasonable results in the interval $8.5 \leq Q^2 \leq 30000$  GeV$^2$, as  shown in Table (1). A shortcoming in these results is  quite small value for the strong coupling constant, suggesting it being kept fixed. For instance, one can uses $\alpha_s(M_Z)=0.119$. This analysis is not presented in the figures.

Concluding the analysis, in  Figs. (\ref{fig2}) and (\ref{fig3}) we verify the   relative role between the hard and soft Pomeron. We present it explicitly for  the virtualities 15, 150 and 8000 GeV$^2$, where the contributions from fitting (I) and (II) are shown. The general feature found is that in the procedure (I), the hard and soft pieces are almost equivalent at small $x$ and intermediate $Q^2$. The soft contribution strongly decreases as the virtualities are large.   From the procedure (II), the soft piece corresponds to a small contribution in contrast with the analysis (I).  In conclusion, we verify that the fitting procedure is equivalent to the model using a two-Pomeron approach \cite{DM,DL}, with the advantage of  clear understanding of the behaviors on $x$ and $Q^2$ of the corresponding hard content and extending our previous analysis made in Ref. \cite{PLB2002}.

\section*{Acknowledgments}

A.L. is grateful to the Organizing Committee, in particular to
Prof. L. Jenkovszky, for the possibility to participate in this  nice
Workshop. M.V.T.M. acknowledges the support of the  High Energy Physics Phenomenology Group at the Physics Institute, UFRGS, Brazil.

\end{document}